\documentclass[aps,10pt,onecolumn,showpacs,pra,citeautoscript,amsmath,amssymb,floatfix,nofootinbib,superscriptaddress,longbibliography]{revtex4-2}

\usepackage[normalem]{ulem}

\usepackage{amsmath}
\usepackage{amssymb}
\usepackage{bbm}
\usepackage{amsthm, thm-restate}

\usepackage{graphicx}

\usepackage{xcolor}

\usepackage{natbib}
\usepackage{hyperref}
\usepackage[capitalise]{cleveref}
\crefname{figure}{Figure}{figures}

\usepackage{algpseudocode}
\usepackage{enumerate}
\usepackage{enumitem}

\definecolor{dgreen}{rgb}{0,.5,.0}

\newcounter{lemmaN}

\newcounter{lemmaA}

\newcounter{colN}

\newtheorem{definition}{Definition}

\newtheorem{postulate}{Postulate}

\makeatletter
\newcommand*{\balancecolsandclearpage}{%
  \close@column@grid
  \clearpage
  \twocolumngrid
}
\makeatother

\makeatletter
\def\tocdepth@fullmunge{%
\let\l@section@saved\l@section
\let\l@section\@gobble@tw@
\let\l@subsection@saved\l@subsection
\let\l@subsection\@gobble@tw@
}%
\def\tocdepth@fullrestore{%
\let\l@section\l@section@saved
\let\l@subsection\l@subsection@saved
}%
\newcommand{\hidetoc}[0]{\addtocontents{toc}{\string\tocdepth@fullmunge}}
\newcommand{\restoretoc}[0]{\addtocontents{toc}{\string\tocdepth@fullrestore}}
\makeatother

\usepackage{mdframed}
\newmdenv[
  backgroundcolor=gray!10,
  linecolor=black,
  linewidth=1pt,
  roundcorner=5pt,
  skipabove=12pt,
  skipbelow=12pt,
  innerleftmargin=10pt,
  innerrightmargin=10pt,
  innertopmargin=10pt,
  innerbottommargin=10pt
]{templatebox}

\usepackage{times} % Sets Times New Roman font
\usepackage{setspace} % Controls line spacing
\usepackage[margin=1in]{geometry} % 1-inch margins

\newcommand{\IQOQI}{Institute for Quantum Optics and Quantum Information,\\ Austrian Academy of Sciences, Boltzmanngasse 3, A-1090 Vienna, Austria}
\newcommand{\Peri}{Perimeter Institute for Theoretical Physics, 31 Caroline Street North, Waterloo, ON N2L 2Y5, Canada}
\newcommand{\VCQ}{Vienna Center for Quantum Science and Technology (VCQ), Faculty of Physics,\\ University of Vienna, Boltzmanngasse 5, A-1090 Vienna, Austria}

\makeatletter
\renewcommand*\l@subsection{\@dottedtocline{1}{1.5em}{2em}}
\makeatother

\begin{document}

\title{Realism about the external world: an adversarial collaboration}

\author{Kelvin J. McQueen}
\affiliation{Philosophy Department and Institute for Quantum Studies, Chapman University, Orange, CA 92866, United States}
\author{Markus P.\ M\"uller}
\affiliation{\IQOQI{}}
\affiliation{\VCQ{}}
\affiliation{\Peri{}}

\date{July 16, 2026}

\begin{abstract}
One of us is a \textit{realist} and believes that reality fundamentally consists in an external physical world, governed by laws of physics. Observers are either emergent (physicalism) or exist in addition to physical reality. The other defends a version of \textit{idealism} and believes that reality fundamentally consists in first-person states, unembedded into worlds, on which laws of nature act directly. Shared ``physical worlds" are merely emergent. The disagreement ultimately centers on the overall coherence of a formal model, known as \textit{Algorithmic Idealism}, and its ability to resolve observer paradoxes, such as the Boltzmann brain paradox. Our aim is to confront this issue in the form of an adversarial collaboration, where we avoid misunderstanding each other, so that readers can see the true source of the disagreement, and decide for themselves. Our work represents an attempt to establish a form of exchange that may help overcome the fragmentation of scholarly communities in philosophy, physics, and elsewhere. The collaboration was conducted using a discipline-neutral template for theoretical adversarial collaboration developed in a companion paper \cite{McQueenMuellerTemplate}, which we hope will be useful in other theoretical disciplines.
\end{abstract}

\maketitle

% Make the TOC more compact so it fits on page 1
\begingroup
\singlespacing        % from setspace; only affects this group
\small                % optional: comment out if not needed
\tableofcontents
\endgroup

\clearpage            % ensure page 2 starts with the introduction

\section{Introduction}\label{Intro}

\subsection{Theoretical adversarial collaboration}

An adversarial collaboration is a research collaboration that aims to resolve disagreement among the collaborators. They are useful in part because they allow the disputants to better understand the nature of their dispute, and to work together towards a solution. This helps avoid the production of critical papers based on misunderstandings, as well as endless objection-reply-rejoinder back and forth often seen in the literature. Adversarial collaborations have become common in social sciences and especially psychology. But elsewhere, we believe they are enormously underappreciated and underutilized. 

We think this is especially the case in our own disciplines: philosophy and the foundations of physics. The relative lack of adversarial collaborations in such disciplines is not surprising. Most adversarial collaborations published to date are \textit{empirical} collaborations, and involve advocates of competing theories working together to construct feasible experiments that distinguish their theories \cite{mellers2001frequency}. Crucially, disputants try to define experimental outcomes that would falsify, or at least reduce their confidence in their theory \cite{corcoran2023accelerating}. But we are interested in resolving disputes that are not so easily resolved by experiment. Such disputes may be resolved by what we will call \textit{theoretical} adversarial collaborations. These are much less common than their empirical counterparts. Without the objective standard of a concrete scientific experiment, it can be much harder to pry a theorist from their theory, raising the question of how fruitful such an adversarial collaboration could be. In fact, in philosophy, we could not find a single journal article that employs a rigorous adversarial strategy.

Here, we present the results of a theoretical adversarial collaboration, conducted in accordance with a discipline-neutral template developed in a companion paper \cite{McQueenMuellerTemplate}. The template is intended for disputes that cannot easily be resolved by experiment. Its condition for success is met when the collaboration yields crucial insights that substantially go beyond existing publications on the theory under dispute: either the critic concedes that the theory can answer the objection through novel and non-obvious arguments, or the advocate concedes that new conceptual resources or revisions are required. In subsequent sections, we explain how our collaboration satisfies this condition.

\subsection{The dispute: realism about the external world}

We will define \textit{external-world realism} (or \textit{realism} for short) to be the view that there exists an external physical world that we all inhabit and that is governed by laws of physics. This view is widely held and is implied by a diverse range of views in philosophy and physics. For example in philosophy, both materialism and dualism imply realism, because both postulate an external material world. They simply differ on whether observers are fundamental too. Materialism denies that observers are fundamental and treats them as emergent. In physics, realism is implied by several interpretations of quantum mechanics, such as Everett many worlds interpretations, Bohmian mechanics, and dynamical collapse theories such as the GRW and CSL interpretations. 

Traditional idealism denies realism by denying that an external world exists at a fundamental level. Observers alone constitute reality, while shared ``physical worlds" are merely emergent, if they are said to exist at all. Although it is a radical view, idealism has a long history in philosophy and has been defended on a variety of grounds. Some have taken idealism to offer a parsimonious solution to the mind-body problem \cite{Chalmers2019-CHAIAT-11}. Others have argued that the external world is redundant, or even meaningless \cite{berkeley1881treatise}. Such arguments were widely discussed in the 18th and 19th centuries. However, the rejection of idealism and development of realism by prominent philosophers at the beginning of the 20th century was profoundly influential and played a key role in the rise of modern analytic philosophy \cite{Hylton1992idealism}.

Why develop and defend a form of idealism? Here we consider a modern descendant of traditional idealism known as \textit{Algorithmic Idealism}, the first version of which was defended in \cite{muller2020law}. Algorithmic idealism was developed to offer solutions to several puzzling thought experiments that have appeared in physics and philosophy. These thought experiments raise difficulties for inferring what the third-person description of the physical world entails for the first person (a difficulty called ``Restriction A'' in \cite{jones_significance_2026}). Here we refer to these difficulties as ``observer paradoxes". A prominent observer paradox (to be discussed in detail below) is the Boltzmann brain problem: if an observer is part of a combinatorially large universe with a huge number of statistical fluctuations (including copies of itself), what should they believe will happen to them? Arguably, it would be a category mistake to try to infer the answer from objective properties of the universe alone. This can be seen as a technical analog of some philosophers' view that a third-person, external perspective on the functional behavior of a human agent can in itself never do justice to the phenomenal internal experience of consciousness, leading to the Hard Problem of Consciousness \cite{Chalmers1996}. 

Furthermore, Bell's theorem of quantum physics undermines certain forms of realism, such as those based on local hidden variables. While this result has motivated many to explore nonlocal realist ontologies, or more exotic ontologies involving ``many worlds", it has motivated others to consider non-realist approaches that emphasize a more agent-centric description of probabilities of measurement outcomes or observed data. Algorithmic Idealism falls into this latter camp and is a principled, mathematically rigorous attempt to resolve all these puzzles together in a simple consistent framework.

Our paper is structured as follows. Section 2 explains what algorithmic idealism is. Section 3 tries to make a case for Algorithmic Idealism, by arguing that it provides a simple and elegant resolution to the Boltzmann brain problem. Section 4 then takes a critical turn, and raises objections to the general framework of Algorithmic Idealism, as well as its attempted resolution of the Boltzmann brain problem. Section 5 then offers responses to these objections, and explains how they have led to the development of Algorithmic Idealism. We leave it to the reader to decide whether algorithmic idealism has successfully met these challenges, with these new developments.

\section{Algorithmic Idealism}

\subsection{Introduction, and conceptual description of the postulates}

In this section, we describe a condensed and slightly updated version of the view presented in \cite{muller2020law,p_muller_algorithmic_2026}. 
Algorithmic Idealism, while it has much in common with traditional idealism, does not fit neatly into traditional categories of ``realism'' or ``idealism'', because it is grounded on notions that are not necessarily related to mind or consciousness. It relies on a notion of ``self states'' (called ``observer states'' in previous publications~\cite{muller2020law}) that are a mathematical notion of abstract patterns or data, and, in this sense, Platonic objects that are neither material nor mental. Algorithmic Idealism assumes that every first person (including you, the reader) can, at every single subjective moment, be exhaustively described as one of those patterns. This intended interpretation, together with the assumption that these patterns are fundamental and do not supervene on a physical world, is a motivation to understand the view as a variant of idealism. 

The self state is interpreted to contain, for example, a description of all your current conscious experiences and unconscious memories; it does not contain, for example, all details of the craters on the dark side of the Moon (unless you happen to have seen them on a photograph). It is thus similar to what some philosophers call a ``person stage''~\cite{Sider1994}. Self states are not in general to be interpreted as human, or as biological, or as in other ways interpretatively familiar. In particular, being a self state does not imply being conscious. For example, you, the reader, are some self state now, but you are also a (different) self state at every moment while you are anaesthesized during a surgery. When Algorithmic Idealism is applied to things we see around us (in a sense to be made precise), then it also applies to structures that are not mental. 

Algorithmic idealism will be defined in terms of two postulates. The first one simply introduces the abstract notion of self states:

\begin{postulate}[Self states]
\label{PostulateSelfStates}
There is a countably-infinite set $\mathcal{S}$ of ``self states'', the collection of patterns that one can be at any given subjective moment. Everything that is to be said about any first person at some moment, including everything there is to be said about its subjective past or future, is determined by its self state.
\end{postulate}

In short, Algorithmic Idealism asserts that you are a pattern, and a discrete one, not a continuous one. Versions of this view are widespread in the Philosophy of Mind~\cite{dennett1991real}, with functionalism as a prominent example. However, most supporters of this view would still make an additional assumption: 
that mental states are physically grounded, or embedded. The received view is that you are not only a pattern, but an \textit{embedded pattern}: you are the pattern \textit{realized on this particular brain in this particular corner of the galaxy of this universe}. You supposedly supervene on some external world in which you are embedded and there is additional self-locating information in addition to your local self state, specifying where and when you are in the world. However, Algorithmic Idealism denies this. Instead, it claims:

\textbf{\textit{You are an unembedded pattern.}}

 This counterintuitive feature is the main novelty of Algorithmic Idealism, enabling new types of resolutions of puzzles such as the Boltzmann brain problem which are unavailable within the received view. Algorithmic Idealism is constructed without initially assuming the existence of a world in which a given self state would be instantiated or realized. Hence it must have other means to ground our apparent ``stream of consciousness'' of self states changing from one moment to the next. The word ``moment'' here, as well as the notions of ``past'' and ``future'' in Postulate~\ref{PostulateSelfStates}, do not refer to any external physical time -- because, by construction, they cannot. Since self states are by definition unembedded, there is no possibility to refer to any physical time, or any time shown on any clock. Furthermore, without assuming a world in the first place, we do not have the usual explanation available for why a given self state was followed by another one --- namely, as a consequence of the evolution of the external world. (Such as: my material brain has evolved in such and such way, which is what caused my self state to change from $x$ to $y$)\footnote{Note that we will soon demonstrate that the availability of an \textit{approximate} explanation of this type is a prediction of Algorithmic Idealism, even though it is not one of its assumptions. In this sense, Algorithmic Idealism makes the same prediction as physicalism in the special case of human observers: the material evolution of the brain will usually exactly determine the first-person experience, as expected. More generally, the evolution of abstract self states will often asymptotically behave as if the self state was embedded into a larger computable probabilistic process, an external world.}. 
Thus, we have to postulate a sort of ``law of nature'' that acts directly on the self state:

\begin{postulate}[State transition]
\label{Postulate2}
If you are in self state $x$ now, you will be in another self state $y$ next, and which one this will be manifests itself as a random experiment for you. Moreover, the likelihood of any $y$ next, given $x$ now, is what would be determined by some universal method of induction.
\end{postulate}

Similarly as Postulate~\ref{PostulateSelfStates}, there are many different ways in which this postulate can be translated into a rigorous mathematical definition. The different ways of doing so give rise to different mathematical models, or implementations, of Algorithmic Idealism.

 Consider your self state $x$ at this very moment. Intuitively, among other things, it contains a description of all your momentary experiences and all the memories collected over your lifetime. This is a \textit{lot} of data. It is not absurd to imagine that a hypothetical external agent performing an ideal method of induction could draw many conclusions from your current self state $x$: for example, that you seem to inhabit a particular planet inside a universe with simple mathematical laws of nature that describe all the physical phenomena that had causal influence on your current state. In particular, this involves many facts that you are not, and never will be, consciously aware of (in particular if you are not a physicist), such as an approximate description of the relevant physical laws.

A method of induction will give us a quantity that we may denote $P(y|x)$, the likelihood of being in self state $y$ next, given that one is in self state $x$ now. This could be a conditional probability distribution, or some other mathematical structure (e.g.\ a quantum state, or a collection of distributions as in imprecise probability, or something else). Given your current self state $x$ with its massive amounts of data, a good method of induction should hence discover the relevant regularities that best explain your current state, and extrapolate them to the future. This should lead to some $P(y|x)$ that is closely aligned with the probabilities that we would infer from our physical laws. Clearly, this intuition must be turned into a rigorous theorem within a concrete mathematical model of Algorithmic Idealism, and we will get to this.

This gives us a hint on how Algorithmic Idealism can be compatible with physics. It also tells us that the expression $P(y|x)$ should not be interpreted as a description of the beliefs that the so-described first person holds, but as objective but private chances. After all, the self described by $x$ may not correspond to anything that would be sentient in any meaningful sense of the word or able to perform calculations; ultimately, $x$ is simply a pattern.

It is important to be aware that several intuitions that one holds from the traditional view (that you are embedded into an external world) have to be dropped. For example, when you have a transition from self state $x$ to $y$, there is fundamentally \textit{no external record} of this.
This is somewhat similar to branchings in the many-worlds interpretation of quantum mechanics. Consider an agent performing a measurement on a quantum state with equal superposition of $0$ or $1$. Everettians understand this as the creation of a unique entangled state, with one branch of the superposition describing an agent that sees $0$ and the other branch describing an agent that sees $1$. From an external perspective, the evolution is deterministic. However, \textit{for the agent}, the situation will manifest itself as a random experiment. After the measurement event, we will have an agent that thinks ``wow, I've seen $0$ and not $1$, and it was completely impossible for me to predict this --- and, in fact, everything that I see around me, and all machinery (including quantum-cryptographic devices) is consistent with the assumption that nobody else could predict it!'' However, there is also an agent (on the other branch) who thinks exactly the same, but with $0$ and $1$ interchanged. That is, there is no matter of fact to the wave function --- no global record --- that would tell us that the outcome of the random experiment was one and not the other. Algorithmic Idealism is similar in this respect: the transitions are assumed to be random experiments \textit{privately for the agent}, but they are not objectively or intersubjectively directly verifiable, even though claims about their likelihood are understood to be objective facts\footnote{However, in the case where we obtain an \textit{approximate} notion of an external world, we will certainly also obtain an \textit{approximate} notion of records of (some) transition events, because it will be possible to understand the transition as a consequence of the physical laws of the resulting emergent external world.}.

\subsection{Preliminary formalization of the postulates: the bit model}

This section introduces a preliminary mathematical implementation of Algorithmic Idealism, called the \textit{bit model}, which has been published in~\cite{muller2020law}. It is not yet a candidate for a final formulation, because it makes a simplifying assumption that is manifestly only approximately correct in some scenarios: that the agent ``fundamentally never forgets anything''. That is, to every given self state $y$, there will always be a unique self state $x$ that was (privately) the case one moment earlier, i.e.\ there is only one possible $x$ such that the transition $x\to y$ is possible. The reason for this assumption is mainly a technical one, since it permits many powerful results from Algorithmic Information Theory to be directly imported (\cite{muller2020law} describes work in progress towards an improved model).

\begin{definition}[Self states]
\label{DefSelfStates}
The set of self states $\mathcal{S}$ is in one-to-one correspondence with the set of finite binary strings,
\[
   \{0,1\}^* = \{\varepsilon,0,1,00,01,10,11,000,\ldots\},
\]
where $\varepsilon$ is the empty string. Given any self state $s$, there are two distinguished self states $s'$ and $s''$, interpreted as the two possible states that the agents may, from his perspective, transition into next. If $s$ is described by the binary string $x_s$, then $s'$ and $s''$ are described by $x_s0$ and $x_s1$, respectively, where the notation indicates the concatenation of the bit $0$ or $1$.

The choice of description by a binary string is not unique: if $s\mapsto x_s$ is a valid choice of description, then so is $s\mapsto \varphi(x_s)$, for every structure-preserving map $\varphi$. These maps $\varphi:\{0,1\}^*\to\{0,1\}^*$ are defined as follows: they are computable bijective maps, with a computable inverse, such that $\varphi(\varepsilon)=\varepsilon$ and $\{\varphi(x0),\varphi(x1)\}=\{\varphi(x)0,\varphi(x)1\}$. All predictions of Algorithmic Idealism will be invariant under the choice of description.
\end{definition}

To understand this somewhat abstract definition, consider a simple example. Naively, we may think of someone's current self state $x$ as a long string of bits that encodes all the functionally relevant details of their brain, up to some level of coarsegraining. But how do we encode the description of this pattern (say, the connectivity of the neurons, and whatever else is relevant) into a finite binary string? In general, there are many ways to do so, and it seems implausible that Algorithmic Idealism would depend on the choice of encoding. For example, if one choice of encoding gives you a string $0011111$, then another encoding can be obtained by simply inverting the bits, $1100000$ (and doing this consistently over time). It seems hard to imagine that one sort of encoding would be privileged over the other, and the two encodings are related by the map $\varphi$ which takes a binary string and inverts it. It turns out that this is a structure-preserving map in the sense of the definition above.

Intuitively, this says that the meaning of a self state $x$ is not given by the succession of bits that describe it, but by the \textit{computability relation} of $x$ to other self states. It is this computability structure on the set of self states that is encoding-invariant. This is somewhat similar to events in general relativity which are described by their coordinates, but all predictions are invariant under diffeomorphisms, and those will in general change the description in terms of coordinates.

Next, we need a mathematical formulation of the state transition postulate:
\begin{definition}[State transition]
\label{DefStateTransition}
If you are in self state $x\in\{0,1\}^*$ now, then you will be in self state $y=x0$ or $y=x1$ next. The private objective chance of this transition is given by conditional algorithmic probability $\mathbf{P}_U(b|x)=\mathbf{P}(xb)/\mathbf{P}(x)$ (defined below), where $U$ is a universal monotone Turing machine. The statistical claims are those that are invariant under the choice of universal machine $U$. (Note that the choice should be \underline{fixed}, and not varied when more and more bits are considered).
\end{definition}

In a nutshell, $\mathbf{P}_U(b|x)$ is large if $xb$ (the concatenation of $x$ with the new bit(s) $b$) is a more compressible extension of $x$, i.e.\ if there is a shorter program that produces $xb$ from $x$, making $b$ a ``more natural'' guess for the future bit(s). We will describe this now in some more formal detail.

A monotone Turing machine $T$ is a particular model of computation, like the standard Turing machine, with a twist: it reads its input bits and produces its output bits one after the other, without erasing anything that it has already written on its output tape, see e.g.~\cite{LiVitanyi}.
We can now imagine a situation in which the input bits are chosen by fair, independent coin tosses, and ask for the resulting probability $\mathbf{M}_T(x)$ that the machine will produce some output that starts with $x$, where $x\in\{0,1\}^*$ is some finite bit string. After producing $x$, the machine may either produce another output bit (and perhaps more), or it may not output anything else. This is why we have $\mathbf{M}_U(x)\geq \mathbf{M}_U(x0)+\mathbf{M}_U(x1)$ (the difference between the left- and right-hand sides is the probability that the machine does not output anything after $x$), and hence $\mathbf{M}_T$ is called a \textit{semimeasure}. The probability measure in Definition~\ref{DefStateTransition} is obtained as the so-called \textit{Solomonoff normalization} of $\mathbf{M}_U$,
\[
   \mathbf{P}_T(\varepsilon):=1,\quad \mathbf{P}_T(xa):=\mathbf{P}_T(x)\cdot\frac{\mathbf{M}_T(xa)}{\sum_b \mathbf{M}_T(xb)}\quad (a\in\{0,1\}).
\]
A monotone Turing machine $U$ is called \textit{universal} if it can emulate every other monotone Turing machine $T$ (for the detailed definition, see~\cite{muller2020law}). If $U$ and $V$ are both universal monotone Turing machines, then $\mathbf{M}_U$ and $\mathbf{M}_V$ are asymptotically close to each other in some sense: there are constants $c,C>0$ such that
\[
   c\cdot \mathbf{M}_U(x)\leq \mathbf{M}_V(x)\leq C\cdot \mathbf{M}_U(x)\quad\mbox{for all }x\in\{0,1\}^*.
\]
This is a surprisingly strong statement. Compare this to a situation where we have, say, a fair and an almost-fair coin toss. The fair coin toss is described by a measure $\mathbf{M}$ with $\mathbf{M}(x)=2^{-\ell(x)}$, where $\ell(x)$ denotes the length of the binary string $x$. On the other hand, an almost-fair coin toss would be desribed, for single bits, by e.g.\ $\mathbf{M}'(0)=\frac 1 2 +\varepsilon$ and $\mathbf{M}'(1)=\frac 1 2 -\varepsilon$. For arbitary strings, we would have $\mathbf{M}'(x)=\left(\frac 1 2 +\varepsilon\right)^{n_1(x)}\left(\frac 1 2 -\varepsilon\right)^{n_0(x)}$, where $n_b(x)$ denotes the number of occurrences of the bit $b$ in $x$. Now $\frac{\mathbf{M}'(1^n)}{\mathbf{M}(1^n)}=\left(1+\frac\varepsilon 2\right)^n$ (where $1^n$ denotes a string of $n$ ones) which tends to infinity exponentially fast, no matter how small the $\varepsilon>0$ is. In contrast, $\mathbf{M}_U$ and $\mathbf{M}_V$ are much less distinguishable than that; in some sense, their difference becomes insignificant for longer and longer strings. For this reason, the (theoretical) scientific practice in algorithmic information theory has been to say that the choice of universal machine $U$ is largely insignificant; the theorems that are proven in this field are always ones that are true \textit{regardless of the choice of the universal machine}. All numerical statements of probability are always up to a global multiplicative constant, for example.

The way that Definition~\ref{DefStateTransition} (and the resulting theory) is supposed to work is along the exact same lines: the predictions of the theory are those that are invariant under the choice of universal machine $U$. These involve in particular predictions for long strings $x$ that describe self states (i.e.\ self states that contain ``a large amount of information''). This is perhaps methodologically comparable to perturbation theory in quantum field theory, where the relevant claims of the theory can only be obtained if some perturbation parameter (say, an interaction strength) is small compared to other physical parameters. There is no mathematical definition of what exactly ``small'' means, but the prescription is nonetheless understood in scientific practice. The practice of algorithmic information theory can serve as a guideline for how to use the claims of the bit model of Algorithmic Idealism for prediction.

In what sense is this an ``inductive'' mathematical structure, as demanded by the conceptual formulation in Postulate~\ref{Postulate2}? It turns out that algorithmic probability $\mathbf{P}=\mathbf{P}_U$ can be used (at least in principle, and in practice approximately) as a method of induction. This is called ``Solomonoff induction''~\cite{HutterBook}: suppose that there is a process that generates a sequence of bits at random, distributed according to some unknown measure $\mu$. We have no idea about this process, except for a promise that $\mu$ is \textit{computable} (which involves almost all distributions that we might be interested in describing or exploring). Suppose that we have so far seen the process output the first $n$ bits, $x=x_1x_2\ldots x_n$. How do we guess $x_{n+1}$, i.e.\ estimate the probability $\mu(x_{n+1}|x)$? The answer: just use $\mathbf{P}(x_{n+1}|x)$ as your guess. In the limit of $n\to\infty$, the difference between this guess and the actual answer will tend to zero, and it will do so optimally quickly in some sense, see e.g.~\cite[Section 1.3.3]{HutterBook}.

Intuitively, the reason why this works is that $\mathbf{M}(x)$ is large for bit strings $x$ that are highly compressible: we have $\mathbf{M}(x)\geq 2^{-\ell(p)}$, where $p$ is the shortest program for the universal machine that makes it output $x$. Consequently, $\mathbf{P}(y|x)$ is larger if and only if $xy$ is a more compressible extension of $x$, i.e.\ one that is a more natural guess for induction (this is true if $y$ is a single bit, or if it is a longer bit string, using the conditional version iteratively). For example, $\mathbf{P}(1|1^n)\to 1$ as $n\to \infty$: if one has seen all 1s and no 0s, then Solomonoff induction predicts that there will probably be another ``1'' next.

Note that Definition~\ref{DefStateTransition} does not claim that there are actual computations going on, or that there are actual monotone Turing machines realized somewhere. The formulation above does \textit{not} say that our universe is a computation. Monotone Turing machines are only used as mathematical ingredients in the definition of the probability measures $\mathbf{P}_U$. As shown in~\cite{muller2020law}, there are alternative ways to characterize these probability measures which do not refer to the notion of a monotone Turing machine.

\subsection{Emergence of an external world and objectivity}
\label{SubsecObjectivity}
In contrast to standard versions of idealism, Algorithmic Idealism does not suffer from an ``external world problem'': despite its non-worldly starting point, it predicts the appearance of an external world. That is, agents (or non-sentient structures) with self states that follow the two postulates of the bit model will probably evolve \textit{as if} they were embedded into a simple, computable, probabilistic external world $W$, with some non-zero probability that depends on $W$. 

In more detail, consider the probability $\mathbf{P}(y|x)$ which determines what happens to any ``self'' according to Definition~\ref{DefStateTransition}, where $x=x_1^n=x_1 x_2\ldots x_n$ consists of $n$ bits. Suppose that we are given some computable probabilistic world $W$ (for definitions see~\cite{muller2020law}) which comes also with a computable pointer that locates some bit string inside of $W$ over time (colloquially, think of a mechanism that tracks your brain inside the galaxy for several years). Then there will be a probability measure $\mu_W$ which describes the chances of what happens to the bit string inside the world $W$. Now the following can be shown:
\begin{equation}
\mbox{With $\mathbf{P}$-probability {\color{dgreen}of} at least }2^{-K(\mu_W)},\mbox{ we have }\lim_{n\to\infty}|\mathbf{P}(y|x_1^n)-\mu_W(y|x_1^n)|=0.
\label{eqEmergenceWorld}
\end{equation}

The terminology in~\cite{muller2020law} calls $W$ a ``computational ontological model'': it is a computable process that can ``explain'' the agent's private state transitions in terms of its temporal evolution, similarly as we think that the evolution of our universe explains our experiences and memories, i.e.\ our self state. Hence, according to Algorithmic Idealism, agents may expect to ``find themselves embedded'' into a computable probabilistic world, at least in the long run --- not in the sense that they \textit{actually are} embedded, but in the sense that \textit{assuming this embedding will faithfully explain their experiences}. Moreover, simpler worlds are more likely: the quantity $\rm{K}(\mu_W)$ corresponds to the shortest program length for any monotone Turing machine that simulates the distribution $\mu_W$ on its output tape (for example, by simulating $W$'s time evolution exactly). If this is small, then $W$ is ``simple'', and $2^{-\rm{K}(\mu_W)}$ is large.

One important further prediction of Algorithmic Idealism is a specific notion of \textit{intersubjectivity}, or, in fact, \textit{emergent objective reality}. To this end, suppose that the event above happens: suppose you are in a self state $x$ such that simple computable probabilistic world $W$ explains what happens to you. Now you \textit{are not actually} embedded into world $W$, but you are \textit{represented} inside of $W$. That is, some part of $W$ (perhaps a biological brain) contains an encoding of your self state. But then, you might look at another bit string inside your world --- say, one that is also growing over time by accumulating more and more bits. For example, you might be interested in the bit string encoded into your friend's brain. Perhaps this is some bit string $x'$; typically, you may know that this string exists within $W$, but you do not know the string. Now you might ask: \textit{what would it be like to be that (unknown) self state $x'$?} Similarly as \textit{your} self state, $x'$ is only \textit{represented} within $W$, but not actually embedded into $W$.

Suppose that you are only interested in a technical version of the above question, namely: \textit{what should the agent corresponding to $x'$ expect to see in the future?} Algorithmic Idealism claims that the private (``first-person'') chances of this agent to transition into future self state $y'$ is $\mathbf{P}(y'|x')$; let us denote this suggestively by $\mathbf{P}_{\rm 1st}(y'|x')$ for now. On the other hand, you may ask: \textit{what will I see my friend seeing in the future?} Since you are pointing at some part of world $W$ (though with a different ``pointer'' than the one that locates \textit{you} inside of $W$), you will use the evolution of your world $W$ to tell you the probabilities for what you will see happening to your friend. This is a distribution $\mathbf{P}_{\rm 3rd}$, corresponding to the ``third-person chances'' of what happens to your friend as a part of world $W$. Now you expect that the first- and third-person probabilities should match: your friend should see what you will see her see! In other words, if you determine (by knowing something about your world $W$) that you will almost surely see your friend see the sun rise tomorrow, then your friend should indeed almost surely see the sun rise tomorrow.

Note that the probabilities here are interpreted as \textit{chances}, not as degrees of belief. Therefore, in our standard view of the physical world, they would trivially be identical. But not so in Algorithmic Idealism: it takes a mathematical theorem with some assumptions to prove that they are close to identical. Namely, it turns out that
\[
\lim_{n\to\infty}|\mathbf{P}_{\rm 1st}(y'|{x'}_1^n)-\mathbf{P}_{\rm 3rd}(y'|{x'}_1^n)|=0.
\]
In fact, $\mathbf{P}_{\rm 1st}(y'|{x'}_1^n)\approx \mathbf{P}_{\rm 3rd}(y'|{x'}_1^n)$ if ${\rm K}(x_1^n)\gg {\rm K}(\mathbf{P}_{\rm 3rd})$: if the complexity of your friend's self state is much larger than the description length of the shortest program that simulates its probabilistic evolution according to world $W$, then your friend and you will ``inhabit the same world'' in a probabilistic sense. In~\cite{muller2020law}, it is argued that we should think of this inequality being satisfied for typical conscious agents that we encounter in our world.

However, the equation above may also be violated, and then we have an instance of an extremely counterintuitive prediction of Algorithmic Idealism: \textit{probabilistic zombies}. These are situations where our usual intuition of objective reality breaks down, such as those described in the context of the Boltzmann brain problem, of some versions of the computer simulation of agents, or death. For a more detailed discussion of probabilistic zombies, see~\cite{muller2020law}.

\section{The case for idealism: solving the observer paradoxes}

\subsection{Overview: when third-person facts do not determine first-person chances}

As an agent or observer, one is often interested in asking the question: ``What happens to me next?'' In standard physicalist accounts, the way to answer this question is according to what we might call the \textit{standard methodology}: first, determine where and what you are in the world (say, this particular material body at this particular place on Earth). Model the evolution of the world (or of its relevant parts) according to the laws of physics. From that, deduce what happens to you. This standard methodology --- or suitable approximations and shortcuts of it --- are successful in all standard situations, including every-day life and all physical experiments that we have constructed so far.

However, this standard methodology is not applicable in all cases of interest. For example, suppose that you are promised to be scanned and reproduced in a computer simulation when you fall asleep next time; perhaps, in fact, a large number of slightly modified copies of you will be simulated, deviating to varying degrees from the ``original you''. Should you expect to ``wake up in the simulation'', and if so, as which copy? Our current physical theories must be silent about this question. You may be desperately longing for guidance -- for example, you might \textit{really} want to obtain a well-justified probability assignment, expressing what you should expect happens to you under these circumstances. However, the physics textbooks will not help you to do this: your candidate probability assignment can never be tested intersubjectively. This is not only a temporary lack of predictability, but a structural in-principle-impossibility: from a third-person point of view (which is the one taken by our current physical theories), there is one copy of you now, and there are many somewhat different copies of you later. This is all there is to say in physics terms, and there is nothing to be uncertain about that would even mandate the use of Probability Theory, or so the argument goes.

Algorithmic Idealism disagrees: it acknowledges that your desperate question \textit{does} make sense. It rejects the idea that the answer to this question would be fundamentally undefined or vague, and it rejects the attitude to say that this question would fall into the realm of the philosophy of mind and be unrelated to physics. Algorithmic Idealism (say, in its bit model implementation) does not quite give you concrete numbers as your probability assignments, but it tells you that you should (ideally, a long time before the exotic situation happens) pick an algorithmic prior $\mathbf{P}$, update it on learning new information, and use it to predict the future, \textit{in both standard and exotic situations}. Or, rather, it tells you that you should think of a hypothetical ideal external rational reasoner doing this that would ideally hold a perfect description of your self state. Even if this is practically impossible, Algorithmic Idealism gives you a useful rule of thumb: private futures that are more compressible given your current state are more likely, also in exotic situations.

Thus, Algorithmic Idealism allows you to obtain private predictions even in cases where the standard methodology does not. This has significant implications for several puzzles of physics and philosophy where the first-person perspective is important. In the next section, we will describe the implications for Cosmology's Boltzmann brain problem.

\subsection{The Boltzmann brain problem}

Let us begin with a very brief summary of the Boltzmann brain problem. Our terminology will mostly follow Carroll~\cite{carroll2020}, and we refer the reader to this paper for more details. A Boltzmann brain (BB) is matter assembled randomly into an exact human brain configuration for a moment, with false memories. In some cosmological models,
BBs are significantly more common than ordinary observers (OOs), if we consider all of time. As Carroll describes it: \textit{``In brief, the BB problem arises if our universe (1) lasts forever (or at least an extraordinarily long time, much longer than $10^{10^{66}}$ years), and (2) undergoes random fluctuations that could potentially create conscious observers.''} In this case, the number of BBs will be vastly larger than the number of OOs. As the argument goes, this means that \textit{we should believe that we are probably BBs}. However, as a consequence, we must conclude that all our memories, including our knowledge of the laws of physics \textit{and all reasoning that led us to this belief}, are probably false too. This leads to what Carroll calls \textit{cognitive instability}: the chain of rational reasoning becomes self-undermining.

This type of argument is more than just a curiosity, because it has a very pragmatic application: if the argumentation above is correct, then we can perhaps use it to constrain our cosmological models. That is, if cosmological model $M$ is consistent with our experimental observations, but predicts a BB-dominated universe, then we might consider $M$ to be ruled out on a theoretical basis. For example, we might argue that science rests (among other things) on the assumption that rational reasoning is possible, and if a physical model undermines this assumption directly, then we should conclude that the model is false. This would give us a very useful tool to rule out cosmological models that would otherwise be consistent with experimental observations.

Realism (or, more concretely, physicalism) enters the argumentation above in an indirect way.  First of all, it suggests that we (or, for concreteness, let us say \textit{you}, the reader) are either one of the OOs, or one of the BBs, but you may not know which one you are. According to this view, we have an instance of self-locating uncertainty: whatever ``you'' are supervenes on some physical stuff, which is embedded into this world (concretely, at some place and time), but you do not know where and when. But this leads to a problem that has also been described in different terminology in two recent preprints~\cite{jones_significance_2026,Adlam}: the third-person perspective of physics does not tell you what belief you should have. Which probability (if any) should you assign to being an OO versus a BB? The designation of a probability value does not correspond to any claim that could be tested intersubjectively by experiment (in contrast to, say, the claim that the Born rule applies in Quantum Theory). One has to invoke principles external to physics to arrive at a formalized degree of belief. One possible candidate is Elga's Principle of Indifference~\cite{elga2004}, which claims that you should assign uniform probability to all OOs and BBs that are locally indistinguishable from you. If you believe that our universe is BB-dominated, you should hence believe that you are probably a BB. This is the reasoning that has been implicitly adopted above.

Algorithmic Idealism rejects the argumentation above and suggests a quite different conclusion. First of all, it rejects the assumption that you are either an OO, or a BB, and you just do not know which one of these is the case. You are neither an OO nor a BB -- you are simply your self state. This self state is \textit{represented} both in all the OOs (that are locally indistinguishable from you) and in all the BBs (that are locally indistinguishable from you), but it would be a category mistake to claim that \textit{you are actually} one of these pieces of matter that represent your self state. It would be a similar mistake as claiming that the number five is actually the number of fingers on your right hand, or it is the number of fingers on the left hand, but we (or it) does not know which one of the two is actually the case.

However, even though the question ``am I a BB or an OO?'' does not make sense within Algorithmic Idealism, a more concrete question \textit{does} make sense: \textit{``Will, whatever happens to me next, be OO-like or BB-like? Which one is more likely?''} Let us try to phrase this within the bit model described above, and keep in mind that this is only the first ``toy model'' attempt to formalize the principles of Algorithmic Idealism. Within this model, only situations are considered where observers learn new information. So, let $x$ be your self state now, and denote the next, say, hundred bits that you will learn by $y$ (so that, after the experiment, your self state will be the concatenation $xy$). Now some of these will describe self states that are typical for OOs (describing, for example, self states where you see ``business as usual on Earth'') -- let us denote the set of all those $y$-strings by $Y_{\rm OO}$. Now, for the BB scenario, let us only consider instances of BBs that do not immediately evaporate, but for which the brain and its associated experiences remain intact for a little while longer. It is clear that the number of so-restricted BBs will typically still outnumber OOs, if Carroll's two conditions above are satisfied. Consider the set $Y_{\rm BB}$ of data $y$ that are typical for BBs. It contains self states corresponding to experiences like \textit{``Wow, suddenly everything is weird and different, and I see all this high-temperature radiation coming in and stuff disintegrating...''}.

Algorithmic Idealism allows you to say something about the conditional probabilities $\mathbf{P}(y|x)$, and it claims in an information-theoretically precise sense that $y$ which are more ``business as usual'' are more likely. Going through the mathematical details provided in~\cite{muller2020law} leads to the conclusion that, indeed,
\[
\sum_{y\in Y_{\rm OO}} \mathbf{P}(y|x)\gg  \sum_{y\in Y_{\rm BB}} \mathbf{P}(y|x);
\]
in this sense, you should expect to make OO-typical observations next, not BB-typical observations. And this conclusion holds \textit{regardless of the number of BBs} that are present in the universe.

In summary, Algorithmic Idealism predicts that it is correct to always prioritize induction over counting: \textit{what happens to you is what universal induction would predict}, and not whatever counting of microstates in conjunction with principles of indifference would predict. Algorithmic Idealism hence solves the BB problem by rejecting the relevance of these thermodynamics intuitions, and it implies that BB counting cannot be used to constrain our cosmological models.

\section{Objections to Algorithmic Idealism}

In what follows I pose two objections to Algorithmic Idealism, the first is aimed at the overall framework, the second at its application to Bolzmann brains.

\subsection{Objection to the overall framework}
\label{SubsecObjectionFramework}

The first objection is that the state transition postulate, and in particular the conditional probability distribution $P(y|x)$, is not well-defined. I will pose a dilemma to argue that this problem is not easily solved, but may instead expose a fundamental problem with the general framework. 

The state transition postulate states that ``If you are in self state \textit{x} now, you will be in another self state \textit{y} next" and it defines the conditional probability $P(y|x)$. The idea is that ``the laws that govern how observers transition from one self state to the next are non-deterministic in that they specify an objective probability $\mathbf{P}(y|x)$ for an observer to transition into state $xy$ given that they are in state $x$". But what exactly are observers or selves, in this model, and how do these objective probabilities attached to them?
Are \textit{selves} fundamental postulates alongside the self \textit{states}? Or are selves emergent somehow, perhaps from patterns of self states? The state transition postulate suggests the former. But the treatment of selves as ``unembedded patterns" suggests the latter. Algorithmic Idealism therefore appears ambiguous between these two options. I will argue that neither option works. 

The two options are very similar to the two theories of the self presented in Parfit's well-known discussion of personal identity \cite{Parfit2009-PARDMA-5}. The first option, where selves are fundamental, is much like what Parfit calls the ``Ego Theory" of self. The second option, where selves emerge from self states, is much like Parfit's ``Bundle Theory" of self. Parfit uses thought experiments to criticize Ego Theory and support Bundle Theory. I will use these same thought experiments to press the dilemma against Algorithmic Idealism. 

The first option for Algorithmic Idealism treats selves as fundamental postulates in addition to the self states. We can then think of the state transitions as describing a kind of \textit{random walk} through the self states \textit{that the observer takes}. Picturesquely, we can use the bit model to imagine all the platonic bit strings, with selves randomly jumping from one bit string to the next. Each bit string is like an empty shell, waiting around until an observer jumps into it. This is like the Ego Theory, where the ego is thought of as a ``soul" which somehow jumps into biological bodies and animates them, until they decease, allowing the souls to then move on. 

This first option, however, is untenable for Algorithmic Idealism. For it makes the number of observers well-defined. There is an infinite number of self states (or bit strings, on the bit model). We can ask whether all of them are inhabited by selves, or whether only some finite number of them are (e.g.\ only bit strings up to a certain size). Either way, there is some well-defined number of them. Yet there being no way to count the number of selves played a crucial role in the application of Algorithmic Idealism to the Bolzmann brain problem. 

Puzzling implications also arise. Imagine you are talking to a close family member. Your experiences of that person in some sense represent the self state of that observer. That observer (from their subjective point of view) has a small chance to transition into a Boltzmann brain state, and experience decaying into thermal equilibrium. Let's assume that (unluckily) they do enter that low probability state. You are left experiencing ``business as usual" but then who (or what) are you left talking to? A mindless self state? A zombie? (This is much like the ``mindless hulk" problem that David Albert raised for single-mind Everettian interpretations \cite{Albert1992}.)

This problem can be made more vivid with a thought experiment that Parfit used to criticize the Ego Theory. Parfit considered split-brain cases, where we are to imagine that after the patient's corpus callosum is split, the left brain is placed and lives on in one body, while the right brain is placed and lives on in another (see also \cite{sider2014personal}). While difficult to execute in practice, this seems like a theoretical possibility that does not violate any physical laws. The ego theorist seems to have to say that the ego or soul jumps to one body rather than the other (insofar as souls cannot bifurcate). What is Algorithmic Idealism supposed to say about this case? We have an \textit{initial} pre-surgery mind $M_i$ that bifurcates into two \textit{successor} minds, $M_{S1}$ and $M_{S2}$. Because $M_{S1} \neq M_{S2}$, it cannot be that $M_i = M_{S1}$ \textit{and} $M_i = M_{S2}$. For Algorithmic Idealism, $M_i$ does not bifurcate into two minds. Instead, $M_i$ has a well-defined algorithmic probability for becoming either $M_{S1}$ or $M_{S2}$: 

\begin{equation}\label{splitbrian}
P(M_{S1}|M_i) \approx P(M_{S2}|M_i) \approx 0.5.
\end{equation}

That is, $M_{S1}$ and $M_{S2}$ are about equally likely future states for $M_i$, because the two sequences of bit strings are about equally compressible.
The first option entails that the observer jumps into one hemisphere, apparently leaving the other uninhabited. Yet, there is no way to tell which is inhabited and which is a self-less zombie! Finally, it is possible (though improbable) that \textit{all} selves jump into one particular self state all at the same time, leaving all but one self state uninhabited - this is difficult to make sense of.

The ``Ego Theory" is useful for making sense of how it could be that the fundamental chances of Algorithmic Idealism attach to \textit{selves}, and give the chances for a self to transition from one state to the next. But it is otherwise implausible and not likely to be taken seriously by advocates of Algorithmic Idealism. Let us then move to the second option, where instead of treating selves as fundamental postulates, we treat them as emergent entities that are vaguely defined but nonetheless supervene on the self states. This is like Parfit's Bundle Theory of self, where selves do not exist in any fundamental way, except that in certain circumstances, perhaps when certain mental states acquire a certain amount of continuity, it is useful to ``bundle" those states together and talk of persistent ``selves". 

Parfit illustrated the Bundle Theory using his ``teletransportation" thought experiments. Imagine I copy all your cells, destroy them, and send a blueprint to a space station, where all your cells are recreated to form a body (and mind) indistinguishable from your own. Have you survived? We can ask the same question after having only recreated a certain portion of your cells, forming a very similar body and mind. The ego theorist seems to \textit{need} to give an answer. But for the bundle theorist there need be no fact of the matter. We know all the fundamental physical facts about what was destroyed and what was created, but they don't conclusively settle when a ``self" survives such experiments. Leading up to teletransportation, there is a clear continuity of mental states, that can vaguely define the persisting self. But fundamentally, there is just the mental states, upon which ``selves" vaguely supervene. Similarly, the bundle theorist has no difficulty with the split-brain case and can allow that there is no fact of the matter as to whether the pre-surgery patient survived. Can Algorithmic Idealism follow the logic of Bundle Theory and treat selves as emergent from self states?

Unfortunately, Parfit's Bundle Theory seems to require realism. In particular, it requires the embedding of mental states into a physical world. A persistent self emerges from a set of mental states when those mental states are embedded in such a way that there is a certain amount of spatiotemporal continuity between them. Such continuity is lost in teletransportation and split-brain cases, so our talk of persistent selves breaks down - but these are extreme cases. Once we have ``unembedded" self states, we lose continuity between states and hence any analogy that might be made with Parfit's Bundle Theory of self. In the bit model, the set of bit strings are not ordered in space and time. They are abstract entities. 

There is also no purely psychological continuity that can be found within the self states, that would help bundle them into a ``self". For example, consider your current self state, which is experiencing this sentence in this paper. There does indeed exist another self state that experiences the next sentence. This might seem like all the psychological continuity between self states one would need. Unfortunately, there is also a self state corresponding to the experience of re-reading the initial sentence instead. There is also a self state corresponding to throwing this paper down and reading something a little less abstract. All these states must exist, and the probability rule assigns them nonzero probability by construction. 

The only other structure in the theory that could help us find emergent selves is the probabilities $P(y|x)$. But it is hard to see how they could help. One might suggest that selves emerge from high probability state transitions. For example, if \textit{x} is most likely to transition into \textit{xy}, and \textit{xy} is most likely to transition into \textit{xyz}, then \textit{xzy} (as opposed \textit{xaa} or \textit{xya}) is a persistent self. But then it becomes impossible by definition for a self to transition into a low probability state. And the split brain case in (\ref{splitbrian}) represents a case where there is no \textit{unique} highest probability state. 

The state transition postulate is therefore not well-defined: there appears to be no good option for spelling out what the probabilities are probabilities of.

\subsection{Objection to the resolution of the Boltzmann Brain problem}

Algorithmic Idealism aims to show that no matter how many BBs are predicted by physics, it is not probable that you are one. In particular, it is not probable that your next self state will be one in which you disintegrate into a maximum entropy environment. Instead, your next state is more likely to be `business as usual'. The argument is that business as usual followed by business as usual is relatively computationally simple, whereas business as usual followed by disintegration is not. The greater the simplicity, the higher the probability, so probably you are not a BB. My objection is to the claim that a BB can actually be described as `business as usual ($x$) followed by disintegration ($y$)'. This is necessary to show that $p(y/x)$ is relatively small. 

How can a BB be described in this framework as initially being in a business as usual state? The answer might seem straightforward: if $x$ only encodes conscious experience (or its neural correlates), then since the BB is momentarily subjectively indistinguishable from an ordinary observer, $x$ captures business as usual. But we know that $x$ must carry more information than that e.g. when you are in dreamless sleep or in a coma, $x$ still has plenty of content - enough to make it probable that you will wake up from a nap. 

But now the problem is that this extra content may encode information about the maximum entropy environment, or at least, its immediate effects on the exterior layers of the BB's cortex. And why would it not contain this information? Indeed an illustrative example in \cite{muller2020law} suggests \textit{it should}. In this example we are to suppose that $x$ describes the state of a little insect that is crawling across the edge of a table. We think there is a large chance $P(y|x)$ of transitioning into a state that experiences falling. Moreover, ``the observer state [self state] should be interpreted as encompassing all information “contained in” the observer, not just what the observer is consciously aware of. In this example, $x$ could contain enough information from the insect’s nervous system to indicate in principle the presence of the table’s edge, even if the insect is not aware of it." If the insect self state contains bodily information indicating a table's edge, then why wouldn't a BB self state also contain information indicating a maximum entropy environment?

If the BB self state inevitably does contain information about its environment, then just as the insect's next ``falling" state is predictable then so too should the BB's next ``disintegration" state be predictable. But then there is no reason to think that the transition probability would be low. But then the framework cannot produce the result that we are probably not Boltzmann brains. 

The only solution seems to be to restrict the definition of the initial `business as usual' BB self state so that it does not include any indication of its environment. But any such restriction seems \textit{ad hoc}, in that instead of the definition being independently motivated, it is defined just so that it solves the BB problem. Of course, insofar as all possible self states exist, then these restricted states must exist. But the states that indicate their environment exist too. And for all you know, you may be one of these broader states that does make being a BB likely.

\section{Idealist responses}

\subsection{Response to the objection to the overall framework}

In summary, the advocate will argue for the following two points:
\begin{itemize}
\item[(1.)] Algorithmic Idealism answers the question of ``what the probabilities are probabilities of'' sufficiently well to obtain concrete predictions for the regime of experience that it targets. Similarly as for Quantum Theory, some aspects of this question are not answered by the theory, but this does not undermine its applicability.
\item[(2.)] There is no fundamental notion of ``self'' in Algorithmic Idealism. The absence of this is a feature, not a bug: in contrast to a Bundle Theory of self, it gives concrete answers to all ``survival questions''; and in contrast to an Ego Theory of self, it does not rely on a dualist or otherwise problematic ontology.
\end{itemize}
The advocate will now elaborate on both points in more detail.\bigskip

\textbf{(1.)} Let us begin with the first question raised in Subsection~\ref{SubsecObjectionFramework}: ``What are these probabilities \textit{probabilities of}?'' Famously, this question applies as well to \textit{Quantum Theory's Born rule probabilities}, and there, it has not found any consensus answer among physicists or philosophers. It is unclear what the quantum probabilities \textit{mean} --- for example, it is an ongoing debate how to relate the Born rule probabilities to the supposed branching structure of an Everettian multiverse. This analogy already suggests the possibility of a pragmatic viewpoint: \textit{whatever these probabilities mean fundamentally}, we understand their operational and pragmatic consequences well enough to obtain concrete empirical predictions for all experiments that can be described within the theory.

My claim is that the same can be said about the probabilities $\mathbf{P}(y|x)$ in Algorithmic Idealism. These probabilities are supposed to express \textit{what you should expect will happen to you next}, under the idealized assumption that you would have a complete description of your current state $x$. Algorithmic Idealism's goal is to answer this question for \textit{private experiments} that are not necessarily intersubjectively verifiable, such as duplication or simulation scenarios. It regards the question of ``what will happen to me next?'' as fundamental and irreducible, and does not attempt to define the notion of ``me'' in this sentence in terms of other notions (more on this in (2.)). In this sense, Algorithmic Idealism is more similar to Operational Quantum Theory (which, similarly, does not define the notion of ``measurement'') than to interpretations like Everettian Quantum Theory (which come with additional metaphysical claims).

While Operational Quantum Theory does not tell you what the quantum state means, it tells you some facts about what it \textit{cannot possibly mean}. For example, due to Bell's Theorem, a local realistic hidden-variable interpretation of Operational Quantum Theory is impossible: the Born rule probabilities cannot simply describe missing knowledge about such variables. Similarly, Algorithmic Idealism tells you, for example, that the Ego Theory of self cannot possibly be applicable (more on this in (2.)). It also tells you that the probabilities $\mathbf{P}(y|x)$ cannot simply be interpreted as describing ignorance about facts of the world, and it tells you that several typical intuitions about probabilities and random experiments must cease to hold.

To illustrate this, and as a preparation for discussing (2.), consider an idealized branching scenario. An agent Alice is duplicated, the two copies make different observations, then every agent is again duplicated, and so forth, for $n$ iterations, where $n$ is large. An external observer would say that there are $2^n$ agents at the end of the experiment. All these agents would claim to be conscious, and would claim to have been Alice at the beginning of the experiment.

Algorithmic Idealism would say that Alice, at the beginning of the experiment, is described by some self state $x$, and there are $2^n$ self states $y_i$, $i=1,\ldots,2^n$, that are represented in the physical bodies at the end of the experiment. According to Algorithmic Idealism, ``what it means to be a self state'' is a property of that self state. In particular, if $x$ and all the $y_i$ are conscious, then we should believe all the agents' claims to be conscious. In particular, Ego Theory does not apply: it is simply not the case that all but one of the agents are ``zombies'' in Chalmers' sense\footnote{The notion of ``probabilistic zombies'' put forward in~\cite{muller2020law} is a completely different concept.}. In fact, none of them are.

Algorithmic Idealism also claims that if you are Alice at the beginning of the experiment, your private objective chance of ending up as any given final agent with self state $y_i$ is $\mathbf{P}(y_i|x)$ . What the above underlines is that there is no matter of fact in the world that any external agent could in the end use to determine ``which event $i$ has actually happened'': formally, $i$ is not a random variable of the probability space that would describe the facts of the world. The meaning of the probability $\mathbf{P}(y|x)$ is about \textit{what you should rationally believe about your private future}, not about an intersubjective future event. In the end, you may look back at what you think you know about your past, and you may assess whether your present corresponds to an outcome that is ``typical'', i.e.\ statistically similar to what you would have expected to obtain, under the claims of Algorithmic Idealism.

Clearly, some of the $2^n$ agents will make \textit{atypical} observations, and they may hence dismiss the initial probability assignment, and consider it as privately empirically rejected. But the point is that you are very unlikely to end up in their shoes. Asking for a sort of infallible confirmation method is arguably a demand that \textit{no probabilistic theory of any kind} can satisfy, including Quantum Theory. You may always have trouble convincing a laymen that winning the lottery is unlikely, if they have just won the lottery, and this does not indicate that Probability Theory is false.\bigskip

\textbf{(2.)} Algorithmic Idealism makes concrete predictions (e.g.\ that you should expect to ``wake up'' in some types of computer simulations but not in others, that you should expect to see a simple computable probabilistic external world, that cosmologists cannot use Boltzmann brain counting to rule out models, and others) without a concept of ``self''. It is simply not needed to make sense of the probabilities $\mathbf{P}(y|x)$ -- the concept of ``self state'' (something similar to a person stage) is sufficient.

The advocate argues that this is a feature of the theory, not a bug. Recall the idealized branching scenario from above. It has already been argued there that the way that Algorithmic Idealism treats aspects of this scenario is in outright contradiction to an Ego Theory of self. Ego Theory would make the problematic assertion that the ``soul'' jumps from $x$ into one of the $y_i$, but no empirical observation could tell an external observer which of the $y_i$ is the conscious one, and which are zombies. Algorithmic Idealism treats all $y_i$ equally, and claims that they are all equally conscious, which seems like a more healthy position to be held.

Similarly, suppose that we rely on a Bundle Theory of self. In this case, the problematic ``soul'' ontology would be omitted, and all the $2^n$ clones could be treated as having spatiotemporal and psychological continuity with the initial agent Alice. However, due to the strong conceptual link with realism, it would be impossible to say anything about the ``ordinary survival'' of Alice. What should Alice expect to observe in the experiment? Physicalism would dictate that the answer to this question should be grounded on facts of the world that can be assessed by external observers, but there are no candidate facts that could play this role.

Algorithmic Idealism, on the other hand, combines the advantages of both approaches while avoiding the disadvantages. It \textit{does} say something about Alice's ordinary survival in the branching scenario, via probabilistic statements on what she should expect will happen to her. At the same time, it avoids the problematic ontology of the Ego Theory of self, by treating all clones essentially equally (even though the transition probabilities $\mathbf{P}(y_i|x)$ will in general be different for different $i$).

\subsection{Response to the objection to the Boltzmann Brain resolution}

It is important to disentangle two different aspects of the critic's objection: first, the question of how BBs relate information-theoretically to their environment; and, second, the consequences of the fact that only part of the self state is consciously available.

Let us begin with some thermodynamic intuition, which generalizes to Algorithmic Information Theory: highly entropic objects have little or no correlations between their parts. The entropy of a composite system $AB$, classically and quantumly, satisfies $S(AB)\leq S(A)+S(B)$, and in order for this to be maximal, we have to have equality. Consequently, the mutual information $I(A:B)=S(A)+S(B)-S(AB)$ is zero: $A$ contains no information about $B$, and vice versa.

Consider some self state $x$, and a cosmological model predicting a huge number of BBs. The vast majority of BB realizations of $x$ will have close to maximal entropy under the constraint of realizing $x$: thus, the BB's state will be almost uncorrelated with its environment. In particular, if the BB remains intact for a while to observe a bunch of new bits $y$ from its environment (as in the bit model), then $x$ and $y$ will be almost algorithmically uncorrelated. This is also intuitively clear: the BB's environment will consist of chaotic fluctuations that are essentially independent of the state of the BB. Thus, the best compression of the new self state $xy$, given $x$, is simply to list $y$ -- no compression at all. This makes $\mathbf{P}(y|x)$ very small. This is different for an ordinary brain in an ordered environment (say, on a planet), where the environment is strongly algorithmically correlated with the brain, making $\mathbf{P}(y|x)$ much larger.

In a nutshell, this is the reason why BB future experiences are much less likely than ordinary ones, \textit{regardless of the number of BBs in the universe}. Consequently, cosmologists cannot use BB counting to rule out some cosmological models. Note that the analysis here gives valid intuition, but not the exact technical arguments: it is not literally the algorithmic mutual information between $x$ and $y$ that is relevant here, but the question of whether there exists a computable probability measure of short program length for which both $x$ and $xy$ are typical outcomes. The technical details are discussed in more depth in~\cite{muller2020law}.

This analysis is independent of another aspect that the critic emphasizes justifiably in their objection: typically, not all of the self state $x$ is consciously available. For the sake of the argument, we can think of $x$ as consisting of two parts, a consciously available part $c$ and an unconscious part $u$, such that $x=(c,u)$ is some encoding of the two strings $c$ and $u$. We can now think of the following two kinds of uneasy cases, where the distinction is not fundamental but one of degree:
\begin{itemize}
\item \textbf{Deceiving self states.} These are cases where rational agents who \textit{only} have access to $c$ (the conscious part) would draw wildly different inferences from agents with full access to $x=(c,u)$. As an example, think of a ``brain in a vat'' scenario: an agent believes to live an ordinary life, while its supposed reality is actually simulated (with observable differences in the future, such as being told what is actually the case).
\item \textbf{Surprisal self states.} Agents with access only to $c$ will in most cases make similar predictions as if they had full access to $x$, but they sometimes become surprised by cases where important unexpected consequences of $u$ show up. The ``crawling insect'' scenario mentioned by the critic is an example of this: according to Algorithmic Idealism, the fact that the insect falls off the table is a property of the emergent external world that is itself a property of its self state $x$, but not in any way that the insect could be consciously aware of (assuming that this category applies to insects at all). It is hence best understood as a property of $u$, and the consequence is that the agent will be surprised.
\end{itemize}
It seems clear that \textit{every} physical theory admitting some notion of observer, including Algorithmic Idealism, admits the in-principle existence of deceiving and surprisal self states. All agents will sometimes be surprised if they lack relevant information about their state, and in essentially all physical backgrounds should ``brain in a vat'' scenarios be conceivable. One may attempt to come up with arguments for why deceiving self state scenarios are practically unlikely, e.g.\ we might think that evolution has largely optimized us for avoiding deep deception of this kind, but such considerations are irrelevant for the present discussion. In particular, Algorithmic Idealism's resolution of the BB problem is \textit{not} based on a claim that deceiving self states would be impossible or a priori unlikely.

The objection by the critic can be rephrased as follows: consider a deceiving self state $x=(c,u)$ encoded in a BB, where $c$ is algorithmically uncorrelated with the BB's environment, but $u$ is significantly correlated with it. In this case, $\mathbf{P}(y|x)$ would be large for a BB experience, even though the agent, knowing only $c$, might want to conclude that ``business as usual'' is their best bet. This argument is correct assuming its premises (that $u$ contains significant algorithmic correlations with the BB's environment). However, the analysis above has shown that this premise is not satisfied: the vast majority of BBs will have all of $x$, and hence also $u$, uncorrelated with the BB's environment. One may even regard this as an information-theoretic definition for distinguishing ``ordinary-planet-like'' realizations from ``BB-like'' realizations: for the latter, there is significant algorithmic correlation between the self state and the environment of its realization, which indicates a common algorithmic cause that is structurally different from a random fluctuation --- a computable mechanism that has generated the agent's state in interaction with its environment.

\subsection{Evaluation of the success of the adversarial collaboration}

We consider our adversarial collaboration a success, in the sense defined in our companion paper \cite{McQueenMuellerTemplate}, because, among other insights, it has brought a substantial difference between physics and philosophy reasoning exactly into the light. A physicist's goal is ultimately to relate their theories to concrete predictions. Algorithmic Idealism was constructed with the same goal in mind, even though it is unusual in aiming for \textit{private predictions}, not ones that can always be tested intersubjectively. It assumes that it makes sense to ask ``What happens to me next?'', without claiming to ground the word ``me'' on other physical or metaphysical notions. It claims that the answer can be formulated in terms of Algorithmic Information Theory, and uses this to derive a couple of predictions, e.g. about what to expect in duplication or simulation scenarios. Comparing predictions like ``the appearance of a simple, computable, probabilistic external world has high probability'' to what we actually see, and observing agreement, is interpreted as evidence for the theory.

The philosopher, on the other hand, scrutinizes the very notions used in the construction of the theory in order to uncover unspoken assumptions, contradictions, or conceptual glitches that may force us to question the theory. For example, a philosopher would like to know what the probabilities $\mathbf{P}(y|x)$ actually mean, even though a complete answer to this question may be irrelevant for many cases of application of the theory. They will want to know what the theory is really supposed to tell us about the world.

Algorithmic Idealism suggests a substantial revision of our usual way to look at the physical world. Hence, even if one takes the point of view that the concrete predictions and not the metaphysical questions should be the focus, technical (physical) and conceptual (philosophical) considerations cannot be considered completely separately from each other. Recognizing the importance of conceptual clarity, our adversarial collaboration has uncovered two valid objections (O1) and (O2). These are not claims that the main postulates of Algorithmic Idealism have to be modified, but that important further work of clarification is left to be done (with some first steps taken in this paper):

\begin{itemize}
    \item[(O1)] \textbf{The interpretation of the probabilities $\mathbf{P}(y|x)$ has to be further elaborated.} One might object that physical theories (broadly construed) may be successful even though the interpretation of their mathematical concepts are not completely clear (think of probabilities in Quantum Theory, for example). However, Algorithmic Idealism is a very unusual approach that cannot (yet) appeal to scientific practice and experimental pragmatism -- for example, in contrast to Quantum Theory or Statistical Mechanics, it is \textit{not} in general possible to relate probability assignments to statistical claims about outcomes of an experiment that is repeated a large number of times under identical conditions, and analyzed from an intersubjective third-person perspective. Thus, the proponent agrees with the critic that this point has to be developed in greater detail and clarity.
    
    \item[(O2)] \textbf{From chance to credence: the self state is not in general fully knowable.} In their objection to Algorithmic Idealism's analysis of the Boltzmann Brain (BB) problem, the critic has pointed at an important fact that has so far insufficiently been discussed in existing publications of Algorithmic Idealism: that only some part of the self state is typically consciously (or otherwise introspectively or operationally) accessible to the associated agent. This raises the general question of how agents should predict their future observations under this epistemic restriction.

In particular, the critic has raised the question of whether this restriction undermines Algorithmic Idealism's analysis of the BB problem. What if some \textit{unconscious} part of the self state contains substantial correlations with the environment (and hence with its future observations) of a BB on which it is realized? The advocate has responded by arguing that the vast majority of BB realizations will be almost uncorrelated with the corresponding environment, both in the conscious and the unconscious part. This is sufficient to confirm that the mere counting of BB realizations does not imply that one should believe that one is a BB, and that cosmologists cannot rule out cosmological models on the basis that they are BB-dominated.

However, the elaboration of this led to an important clarification of a misunderstanding that has been falsely raised against Algorithmic Idealism in the past: the idea that Algorithmic Idealism would predict that every agent's futures are maximally unsurprising, i.e.\ ``boring'', because what happens with high probability is what would likely be predicted by universal induction. However, even fully deterministic and algorithmically simple evolutions which are hence perfectly and deterministically predicted by universal induction can lead to extremely surprising behavior, as demonstrated for example by cellular automata like the ones famously popularized by Wolfram~\cite{Wolfram}. Moreover, if this objection were correct then it would also have to apply to standard physics, and it does not, because humans not being bored is no evidence against the validity of the usual physical laws. Delineating the possibility of surprisal from the impossibility of BB-like ``cognitive instability''~\cite{carroll2020} has hence emerged from this adversarial collaboration as an interesting direction for further research.
\end{itemize}

We believe that progress lies in the combination of the predictive and the conceptual viewpoints, and that adversarial collaborations between physicists and philosophers can be a fruitful tool to achieve this. The advocate certainly feels that they have benefited immensely from this collaboration, and that they understand Algorithmic Idealism now much better than before.

\section*{Acknowledgments}

This research was supported by grant number FQXi-RFP-CPW-2015 from the Foundational Questions Institute (FQxI) and Fetzer Franklin Fund, a donor advised fund of Silicon Valley Community Foundation.

This research was supported in part by Perimeter Institute for Theoretical Physics. Research at Perimeter Institute is supported by the Government of Canada through the Department of Innovation, Science, and Economic Development, and by the Province of Ontario through the Ministry of Colleges and Universities.

\bibliography{bibliography}

\end{document}